\documentclass{jaa}
\usepackage{natbib}
\usepackage{color}
\usepackage{dirtytalk}
\usepackage{graphicx}
\usepackage{multirow}
\newcommand{\astrosat}{{\it AstroSat}\,\,}
\newcommand{\jude}{{\it JUDE}\,\,}

\newcommand{\bit}[1]{\ensuremath{\textit{\bfseries{#1}}}}
\def\barr{\begin{array}}
\def\earr{\end{array}}
\def\berr{\begin{eqnarray}}
\def\err{\end{eqnarray}}
\def\berrno{\begin{eqnarray*}}
\def\errno{\end{eqnarray*}}

\begin{document}\sloppy
\definecolor{lightgrey}{gray}{0.80}

\newcommand{\cmdshell}[1]{\vspace{0.3cm} 
        \colorbox{lightgrey}{\makebox[\textwidth][l]{#1}}
        \vspace{0.3cm}
}
\newcommand{\cmdlist}[1]{\vspace{0.3cm}\centerline{
        \framebox[\textwidth][l]{\small{#1}}}
}

\title{JUDE (Jayant's UVIT Data Explorer) Pipeline User Manual}


\author{P T Rahna\textsuperscript{1*,2}, Jayant Murthy\textsuperscript{1} and Margarita Safonova\textsuperscript{1}}
\affilOne{\textsuperscript{1}Indian Institute of Astrophysics, Bengaluru,560034, India.\\}
\affilTwo{\textsuperscript{2}CAS Key Laboratory for Research in Galaxies and Cosmology, Shanghai Astronomical Observatory, Shanghai, 200030, China.}


\twocolumn[{

\maketitle

\corres{7rehanrenzin@gmail.com}

\msinfo{28 Oct 2020}{13 Dec 2020}

\begin{abstract}
We have written a reference manual to use \jude (Jayant's UVIT data Explorer) data pipeline software for processing and reducing the Ultraviolet Imaging Telescope (UVIT) Level~1 data into event lists and images -- Level~2 data. The JUDE pipeline is written in the GNU Data Language (GDL) and released as an open-source which may be freely used and modified. GDL was chosen because it is an interpreted language allowing interactive analysis of data; thus in the pipeline, each step can be checked and run interactively. This manual is intended as a guide to data reduction and calibration for the users of the UVIT data.
\end{abstract}


}]


\doinum{12.3456/s78910-011-012-3}
\artcitid{\#\#\#\#}
\volnum{000}
\year{0000}
\pgrange{1--}
\setcounter{page}{1}
\lp{1}

\section{Introduction}

This manual describes the \jude software system for processing and understanding the Ultraviolet Imaging Telescope (UVIT) data. UVIT was launched on 28 September 2015 on a PSLV-C30 into a 650 km,  $6^{\circ}$ inclination low Earth orbit by ISRO (Indian Space Research Organization), as part of the multi-wavelength Indian \astrosat mission. The UVIT has seen its first light on 30 November 2015 by observing the  open cluster NGC188. UVIT operates in three channels: visible, near-ultraviolet (NUV), and far-ultraviolet (FUV), covering a broad range of UV to the visible spectral region from about 1250 to 5500 \AA, dividing it into 15 narrow and broad wavelength bands.

The UVIT instrument has been continuously observing since 2015, and the data, including the PV (performance verification) phase observations used to characterize the instrument \citep{Tandon2016} with in-flight calibration and verification \citep{{Rahna2017},{Tandon2017}}, are being released to the observers. The Announcement of Opportunity (AO) Cycle was initiated in April 2017 for the Indian astronomy community. Though at present (2020) only the FUV and VIS channels are operating (NUV channel has failed in March 2018), the archive is large and open to the public --- more than 1000 datasets of UVIT data were made public. The data may be accessed at the \astrosat data archive\footnote{https://astrobrowse.issdc.gov.in/astro\_archive/archive/Home.jsp --- one needs to register and create an account there.} (Astrobrowse) at the Indian Space Science Data Centre (ISSDC), Bangalore. Further data will be released upon completion of the respective proprietary periods (12 months). \astrosat has now completed its 5 years of operation.

The official UVIT pipeline is developed and available for download at the Science Support Cell ({\small \tt http://astrosat-ssc.iucaa.in}). It is designed to produce scientifically usable images from the raw data -- Level~2 data, which are flat-field, distortion, and drift-corrected images with absolute calibration. Canadian UVIT pipeline called CCDLAB is also available for UVIT data reduction \citep{2017postmapipeline}. However, the source codes of these two pipelines are proprietary and difficult to modify. 

We have written the software package \jude (Jayant's UVIT data Explorer) entirely in the GNU Data Language (GDL) \citep{GDL2010} to reduce Level~1 UVIT data into Level~2 image files and event files. \jude is released under an Apache License\footnote{http://www.apache.org/licenses/LICENSE-2.0}, which may be freely used and modified. GDL was chosen for the development environment because it is an interpreted language which lends itself to the interactive analysis of data. This was invaluable in the development of the pipeline, where we could check each step and run commands interactively. GDL is an open-source version of the Interactive Data Language\footnote{https://idlastro.gsfc.nasa.gov/} (IDL) and runs on a wide variety of systems, just as IDL. GDL will run most IDL programs without modification (and {\it vice versa}), allowing access to the rich library of utilities developed for IDL over the last four decades, thus easing the development of \jude. We have tested \jude using both GDL (version 0.9.6) and IDL with identical results on multiple operating systems. In the remainder of the manual, GDL and IDL may be freely interchanged. \jude starts with the Level~1 data, provided by the Indian Space Science Data Centre (ISSDC), and produces photon lists and images suitable for the scientific analysis. \jude is archived at the Astrophysics Source Code Library \citep{Murthy_ascl}, and the latest version is available on GitHub at: {\small \tt https://github.com/jaymurthy/JUDE}.

We will continue to update the pipeline as we find errors and also welcome feature requests to improve the scientific utility of the programs\footnote{As was communicated to us by a user, the use of a /notime keyword in the \textit{interactive.com} code was giving an error message about exposure times. We have corrected this in the code; Nov. 2020.}.

\subsection{UVIT Instrument details}

\begin{table*}
\caption{Summary of characteristics of UVIT instrument}
\label{table:UVIT}
\begin{tabular}{|l|l|}
\hline

Parameter & Value \\ \hline
Type$^{a}$        & Ritchey-Chr\'etien       \\ 
Diameter$^{a}$        & 37.5 cm       \\
Field of View $^{a}$          & 28$^{\prime}$       \\ 
Detector$^{a}$      & CMOS, 512 $\times$ 512      \\ 
Mode of operation$^{a}$      & Photon counting mode (FUV and NUV)     \\ 
Channels$^{a}$        &  FUV ($1300-1800$ \AA), NUV ($2000-3000$ \AA)        \\
Spatial Resolution$^{b}$        & $\lesssim 1.5 ^{\prime\prime}$        \\ 
Frame rate$^{a}$         & 29 frames/second     \\ 
Sensitivity$^{a}$            & 20 AB Mag in FUV for 200 s        \\ 
Time resolution$^{c}$            & $<$ 35 ms      \\ 
Pixel Scale             & 3.28$^{\prime\prime}$ (for 512 $\times$ 512 window size)      \\ 
\jude pipeline image format  &  4096 $\times$ 4096 pixels with 0.41$^{\prime\prime}$ pixel size   \\  \hline
\multicolumn{2}{l}{$^{a}$ \cite{Kumar2012}, $^{b}$ \cite{Rahna2017},  $^{c}$ \cite{murthy2017jude}}

\end{tabular}
\end{table*}

The UVIT is one of the five payloads on \astrosat spacecraft. It consists of two identical 37.5 cm Ritchey-Chr\'etien telescopes, one is reserved for FUV channel, and the second one sharing NUV and visible (VIS) channels through a beam splitter (Table~\ref{table:UVIT}). The VIS channel is not intended for science purposes and is used solely to track the spacecraft (s/c) motion. Each channel is equipped with a 5-filter filter wheel, providing spectral coverage in several passbands from the FUV to the visible. The basic properties of UVIT filters are given in Table~\ref{filterprop}. The effective bandwidth $\Delta \lambda$ in Table~\ref{filterprop} is the integral of the normalized effective area, and $\lambda_0$ is the central (or `mean') source-independent wavelength, weighted by the normalized effective areas; these quantities were calculated from the ground calibration values \citep{Rahna2017}. See \cite{Tandon_2020} for the latest updated corrections to the ground calibration values.

\begin{table*}[h!]
\centering
\caption{Properties of different UVIT UV filters.}
\label{filterprop}
\begin{tabular}{cclcccccc}
\hline 
Filter  & Type & Filter name & Passband & Effective bandwidth  & Central $\lambda_0$ & CF$^{a}$ & ZP$^{b}$   \\
Channel &  &  & (nm)&  $\Delta\lambda$ (nm) &  (nm)  &  (erg cm$^{-2}$ \AA$^{-1}$ {\rm cnt}$^{-1}$) &  (mag) \\     
\hline
\multicolumn{1}{c}{\multirow{6}{*}{FUV}}  & CaF$_{2}$-1  & F148W&  $125 - 179$ & 44.1  & 150.94  &  3.8689e-15 & 17.73    \\
\multicolumn{1}{c}{}  & BaF$_{2}$  &F154W& $133 - 183$ & 37.8  &  154.96 & 4.2036e-15 &17.58  \\
\multicolumn{1}{c}{}   & Sapphire  &F169M& $145 - 181$ & 27.4&  160.7  & 5.4399e-15 &17.22 \\

\multicolumn{1}{c}{}   & Silica    &F172M&  $160 - 179$ & 13.13&  170.3  &  1.4273e-14 &16.05 \\
\hline
 \multirow{6}{*}{NUV}& Silica           &N242W& $194 - 304$ &  76.9   &   241.8         & 1.9459e-16 &19.95 \\
 & NUV15           &N219M&  $190 - 240$ & 27.1       & 218.5     &  5.8360e-15 &16.48 \\
 & NUV13           &N245M&  $220 - 265$ &  28.17     &    243.6        & 1.0327e-15 &18.12  \\
 & NUVB4           &N263M&  $220 - 265$ &  28.23      &   262.8   & 6.9611e-16 &18.39  \\
 & NUVN2           &N279N&  $273 - 288$ & 8.95       &   279.0  & 2.7988e-15 &16.75  \\
\hline
\multicolumn{8}{l}{$^{a}$ Conversion factors from \cite{Rahna2017}} \\
\multicolumn{8}{l}{$^{b}$ Zero-point magnitudes derived from conversion factors } 

\end{tabular}
\end{table*}

\subsection{Status of \jude}

\begin{itemize}
\item The development of \jude began in June 2016
\item The production release of \jude is June 2017
\item \jude\! update: June 2019

\item The last update of \jude\!: November 2020

\end{itemize}

\section{UVIT Data Definitions}

\subsection{Level~0 data}

Level~0 data is the raw data from the \astrosat spacecraft. This data is sent to the \astrosat Data Center at the ISSDC, where the data is separated by instrument and written into Level~1 data.

\subsection{Level~1 data}

The Level~1 data from ISSDC is sent to the payload operations center (POC) at the Indian Institute of Astrophysics (IIA), Bengaluru. The Level~1 data are distributed as a single zipped archive for each observation. The files within the archive are in a number of sub-directories organised by the orbit number and the type of file, all under a single top-level directory. All data files are FITS binary tables. More details about the data are described in \cite{murthy2017jude}.

\subsection{Level~2 data}

Once the refined coordinate information is brought into the FITS binary extension tables, it is ready to be converted to an image of processed data from a single scan of the sky. These images are termed Level~2 data. This is the final data product from the pipeline software. Level~2 data are FITS images, readable by any of the standard astronomical data processing packages. 

\section{Installing the \bit{JUDE} Pipeline}
\label{chapt:judeinsta}
\subsection{Setting up \jude}

\begin{enumerate}
\item \textbf{Download and install GDL/IDL:} \jude runs under either IDL or GDL. You have to download and install either of them to run \jude software system.
\item \textbf{Download the IDL libraries:}
\jude uses many library routines from the IDL Astronomy Users' Library\footnote{https://idlastro.gsfc.nasa.gov/ftp/astron.zip} 
and from the Markwardt IDL Library\footnote{http://www.physics.wisc.edu/\(\sim \)craigm/idl/down/cmtotal.zip}  
that can be downloaded and installed. These routines should be placed in one of the directories pointed to by the !PATH variable. Our recommendation is to create a separate directory for each library and unzip the files into that directory. Update the !PATH system variable to include these directories.
\item {\bf Download \bit{JUDE} programs from GitHub.}\footnote{https://github.com/jaymurthy/JUDE}
Add the \jude  library to the IDL path i.e,: !path = path to jude/: + !path
\item \textbf{Download Level~1 data of your target from ISSDC Astrobrowse.} Unzip and save the downloaded raw data in a folder called Level1(dname). Note that \jude will find all relevant files under the dname directory. The individual files may be gzipped to save considerable space.
\end{enumerate}

\subsubsection{{Running \jude Pipeline}}

\subsection{Automatic mode}

\jude comprises a number of individual routines that together can be run as a pipeline. A program ({\it process\_uvit.com}) is included in the \jude distribution, which may be modified for local use and can be run using the syntax:

\noindent
\textbf{Step 1: \textit{@process\_uvit.com} (from IDL)
or \textit{idl process\_uvit.com} (from terminal)}.
\\
\noindent
This program chains together the following commands (Note that, if desired, these statements may be entered in sequence at the GDL prompt). You may have to set the path before running \textit{process\_uvit.com} :
\begin{enumerate}
\item 
Setting path (as per above setup):
\begin{itemize}
\item $!path$ = ``$<$ path to jude folder $>$ :''$ + !path$; Add \jude  routines to path.
\item $ !path$ = ``$<$ path to  cmlib folder $>$
 :'' +$ !path $; Add Markwardt routines.
\item $!path$ = ``$<$ path to idl libraries $>$ :'' $+ !path $; Add IDLASTRO routines.
\item dname = ``$<$ path to level1 folder $>$''; Location of Level~1 data.
\end{itemize}
\item Produce Level~2 data:
\begin{itemize}
\item jude\_driver\_vis, dname; 
\item jude\_driver\_uv, dname,/nuv,/notime
\item jude\_driver\_uv, dname,/fuv,/notime
\item jude\_verify\_files, dname
\item jude\_uv\_cleanup, /nuv
\item jude\_uv\_cleanup, /fuv
\end{itemize}
\end{enumerate}
{\bf Description}: \textit{jude\_driver\_vis} will process the visible files (VIS) and \textit{jude\_driver\_uv} will process the UV files to produce event lists and images. When running \textit{jude\_uv\_driver} for the first time, it is faster to use the /notime keyword. The time per pixel will be approximate at the edges. Program \textit{jude\_verify} was introduced because there are crashes in GDL for long runs, and it will go through each of the files and ensure that they are, at least, readable. \textit{jude\_uv\_cleanup} will process the event lists and images, merge the data, and register the images. Finally, it creates Level~2 data.\\
There are random crashes caused by memory issues which can affect datasets with many files. These errors may occur in any module and at different locations. It was found that 4 GB RAM is enough but, obviously, more RAM is better. In most cases, the program may be restarted and will pick up from where it was left off, skipping files already processed. If the memory problem occurs at a critical point, it is possible that the FITS file may not be properly written. We have added a program (\textit{jude\_verify\_files}) which should check the integrity of each FITS file and delete it if there is a problem. If this does not work, it would be best to delete any suspect files and start \textit{process\_uvit.com} again. Files already completed will be skipped. If the verification (from \textit{jude\_verify\_files}) is successful, a file named \textit{jude\_VERIFY\_FILES\_DONE} will be created. There will be no further Level~1 processing if this file exists. It should be deleted if processing from the scratch is really desired.

\subsection{Interactive mode}

After the initial run of \jude using \textit{process\_uvit.com}, you can run it interactively with \textit{interactive.com} using the syntax:
 
\noindent 
\textbf{Step 2. \textit{@interactive.com} (from IDL) or \textit{idl interactive.com} (from terminal).}

\noindent
It will open a window with the image and ask a few questions:
\begin{itemize}
\item  Enter new image scaling (Enter if ok)?

If image is too messy or crowded, you can change the contrast (scale), e.g. 0.001. 
\item  Parameter to change?
\say{-1} to continue, \say{-2} to exit, \say{-3} to debug: 

It will display default parameters on the screen: The explanation of these parameters is given in Table~C.5. in the appendix to the \jude paper (Murthy et al. 2016). You can change the parameters based on your data by entering the corresponding parameter number and change the values. If you don't want to change the default parameters, type \say{-1} and hit Enter.

\item  Run centroid (y/n/v/r)? (\say{y} -- yes, \say{n} -- no, \say{v} to use VIS, \say{r} to reset offsets):
\begin{itemize}
\item If you want it to run automatically using \jude centroid program, press \say{y} and Enter.

\item If you want to correct for s/c motion using VIS data interactively, press \say{v} and Enter.

\item If you choose \say{n}, it will run with default offset values and gives you the final output with default s/c  correction.

\item If you press \say{r}, it will reset offsets (make them full nulls).
\end{itemize}
\item  Run using \jude centroid:
First, it will ask to change the image scaling, then it will ask to choose the individual star from the displayed image. Select the isolated bright star in the field within the binned image by clicking the cursor on the display image. It is better to avoid stars on the edges of the field.

\item `Enter new box size (or press Enter):' You can specify a box size here, and the program will perform the s/c correction by following the selected star within the box size in each set of frames, and display the co-added image. It will also display the PSF of the image.

\item `Write files out (this may take some time)?': If you are satisfied with the PSF and the image, you can write the FITS file by pressing \say{y} to this question.

\item `Do you want to run with different parameters?': If you want to improve the images, you can run the code again by entering \say{y}, otherwise it will automatically go to the next image. Follow the same procedure.
\end{itemize}
\noindent
After finishing all NUV, it automatically goes to FUV. Follow the same procedure.

\noindent
\textbf{Important note:}  When you run {\it interactive.com}, you can use /notime keyword with \textit{jude\_interactive.pro} in \textit{interactive.com} to run fast. But, in order to get the correct exposure time and flux (CPS) in the final output, remove the /notime keyword from {\it interactive.com} and run it again, or run \textit{jude\_apply\_time.pro} afterwards for all the files.  \\
\textbf{Calling sequence: \textit{jude\_apply\_time, L2\_data\_file, uv\_base\_dir, params = params}}\\
where {\bit{L2\_data\_file}} is the directory of Level~2 events list folder (/events) and {\bit{uv\_base\_dir}} is The top level UV directory (/fuv or /nuv).

\section{Data Products}

\begin{itemize}
\item \textbf{Events list:} A FITS binary table with $N$ extensions, where $N$ is the number of frames in the data. Each extension has a list of all the photons detected in that frame. The format of the file is described here: in the first stage of processing, one event file will be created for each Level~1 data file. Duplicate files will be removed in the second stage (\textit{jude\_uv\_cleanup}) and the data will be merged if they are continuous. There will thus be fewer event files than Level~1 data files.

\textbf{Display events:} \textit{make\_movie.pro} will provide a movie of the frames from the events files, which can be run using \textit{make\_movie, events\_file}.

\item \textbf{Images:} A FITS image file will be created for every event file. Each frame will be shifted and added to correct for spacecraft motion.\\
The first extension in the FITS image file will be the flux in counts/second (CPS).\\
The second extension in the FITS image file will be the exposure time for each pixel in seconds. If /notime keyword was set, then the array contains the total number of frames. 
The final image from \jude will be $4096\, \times\, 4096$ pixels ($28^{\prime}\times 28^{\prime}$) with the pixel size of 0.41$^{\prime\prime}$. The total exposure time will be given in the fits header of each images.

{\bf Changing exposure time:}
The exposure time of a final image can be changed by changing parameters ``minframe'' and ``maxframe'' when running {\it interactive.com}. This can be helpful if the user is looking for flux variability at different time scales.

\item \textbf{Diagnostic files:} Corresponding quick-look PNG image will be created for every event file. This image will additionally contain the following plots: a plot of the data quality index (DQI) for each frame, a histogram of the number of counts per frame, and a plot of spacecraft motion derived from calculating shifts between the frames in both $X$ and $Y$. For details on DQI and  calculation of shifts see \citet{murthy2017jude}.
\item \textbf{CSV files:}\\
{\it final\_obslog.csv}: Observation log for Level~2 data;\\
{\it obs\_times.csv}: Observation log for Level~1 data with duplications;\\
{\it merge.csv}: Log of files merged to produce the final Level~2 data.
After running \textit{interactive.com}, \textit{jude\_obs\_log.pro} can be run to create the observation log of final images. \\ \textbf{Calling sequence: \textit{jude\_obs\_log, data\_dir, output\_file, params.}}

\end{itemize}

\section{Verifying the Output}

In principle, \jude should automatically produce usable science products: photon lists and images. However, there are times when the registration fails, and it is best to verify the images manually. Program {\it interactive.com} will run through each Level~2 file and give options for reprocessing the data, if desired. The main purpose of the program is to run \textit{jude\_interactive.pro} for each event file. As a general rule, there are three parameters to check:
\begin{itemize}
\item The integration time of the image should be close to that of events file. If it is much less than the actual time observed, it is worth checking why. 
\item The PSF depends on the quality of the correction for the s/c motion, and is typically 1.2--1.5 arcseconds. If the PSF is much greater than that, it means that automatic registration failed. 
\item The appearance of the image depends on the s/c motion, and this may be seen as tails in the sources.
\end{itemize}
\noindent
\textit{jude\_checkpsf.pro} can be used to check for the PSF of a star in the field. The program gives PSF in $x$ and $y$ direction (unit of both pixels and arcsec) and also displays 3D surface profile and a contour plot of the star. \\
\noindent
\textbf{Calling sequence: \textit{jude\_checkpsf, im, res}}\\ where {\bit{im}} is the Level2 image file of UVIT, and {\bit{res}} is the resolution (number of sub-pixels in one physical pixel) of the image. By default, \jude produce images with 4096 $\times$ 4096 (res=8) pixels. 

\section{Data Registration}

Data registration corrects for spacecraft motion. The first estimate comes from the VIS images. The program \textit{jude\_match\_vis\_offsets.pro} only samples every 1 second in time, and thus the derived spatial resolution is poorer. The second estimate comes from self registration -- program \textit{jude\_centroid.pro}. As long as there is a bright star in the UV channel, this will give better resolution than getting offsets from the VIS channel. Most fields have stars that are bright enough to follow along. If the automated registration fails, there are several steps to follow: 
\begin{itemize}
\item Increase the number of bins (the amount of time) to be binned over. The default is 20 frames (0.6 seconds). The box size usually has to be increased because the s/c motion is larger in longer times.
\item Decrease the resolution, which effectively increases the S/N per pixel. It is usually easier to find bright sources suitable for registration in the NUV channel. We have written {\it jude\_match\_nuv\_offsets} code to match the NUV offsets to the FUV channel.
\end{itemize}

\section{Astrometric Corrections}

We have found that the astrometric information provided by the s/c star sensor is not accurate enough for good astrometric calibration of the UVIT images, and we have implemented several alternatives. 

\subsection{Astrometry.net}

This is our recommendation and it gives excellent solutions for most NUV fields. Though it is less likely to work well for the FUV files because the stellar magnitudes are much more different from the optical. There are several different ways to use Astrometry.net. 

\subsubsection{Directly on the website}

\begin{itemize}
\item Submit images to the website (http://nova.astrometry.net). Anonymous uploads (without signing in) are allowed but then the data becomes public. Creating an account allows more customization.
\item The solution will be faster if we restrict the field size to the UVIT FOV.
\item The FITS file with the astrometric solution will be on the ``SUCCESS" page and is named \say{new-image.fits}.
\item Note that this file will take the first extension from the original file (the data) and will not include the exposure time per pixel.
\item We have written a program  (\textit{jude\_copy\_astrometry\_hdr.pro}) to merge the astrometry from the new-image.fits into the original file.\\
\textbf{Calling sequence: \textit{jude\_copy\_astrometry\_hd, astrometric, l2\_file}}\\
\noindent 

\end{itemize}

\subsubsection{Python script to upload data files to Astrometry.net}

\begin{itemize}
\item Python and client.py are required to run this.
Typing \say{python client.py} will print the arguments to the program.
\item A key (API\_KEY) to Astrometry.net (obtained through creating an account) is also required.
\item \textit{jude\_call\_client.py} is the interface to client.py.
The key may be stored in the system variable  AN\_API\_KEY; entered as a keyword, or entered when requested.
The path to client.py should be changed in the program (if it doesn't exist, the user is prompted for input).
The client.py is called with preselected parameters. The Level~2 image is uploaded to the Astrometry.net website, processed online, and then downloaded. Finally, We update the astrometry in the Level~2 file.
The call would be (with default parameters) \textbf{\textit{jude\_call\_client\_py,inp\_dir,out\_dir}}. {\bit{inp\_dir}} is the directory containing image files to be corrected and {\bit{out\_dir}} is the directory to put files from astrometry.net.
\end{itemize}

\subsection{Independent astrometric corrections}

There are times when the astrometric correction fails, more often in the case of the FUV where there are not enough recognizable stars. Program \textit{jude\_fix\_astrometry} takes an image with known astrometry (typically the NUV image) and compares it with the test image. We have tried to automate the procedure as much as possible, but user intervention is necessary at times. At least two points are required for an astrometric solution. If two or three points have been defined, the code {\it starast.pro} is used. If 6 or more points have been defined, the code {\it solve\_astro.pro} is used.  In case not enough points are found, they can be added by hand to a reference file. The name of the reference file is the same as the image file with \say{.FITS.gz} replaced by \say{.ref}. The reference file has 4 lines, each with at least two elements. The lines contain the $x$ and $y$ positions (in pixels) of each star with its RA and DEC (in degrees). The order of the lines is unimportant, but the variable names ($newxp, newyp, newra, newdec$) are required. e.g.,
\berrno
&&newxp=[3494.0319, 3099.7382, 1987.0005]\\
&&newyp=[1579.9278, 1483.6838, 2929.903]\\
&&newra =[9.1011347, 9.1398875, 9.4078713]\\
&&newdec=[39.927001, 39.963891, 39.913863]
\errno

\textbf{Calling sequence: \textit{jude\_fix\_astrometry, new\_file, ref\_file}},\\
\noindent 
where {\bit{new\_file}} is the Level~2 image file to be corrected astrometrically, and {\bit{ref\_file}} is file with accurate astrometry.

\subsection{Coordinates for every photon}

\textit{jude\_apply\_astrometry.pro} will take an astrometrically corrected image file and apply the astrometry back to the event list. The $x$ and $y$ of each photon (in the detector plane) will have an associated RA and DEC.

\section{Co-addition of images}

The final step is the co-addition of the images from different orbits. We used \textit{jude\_coadd.pro} routine to produce final co-added image. Program \textit{jude\_coadd.pro} will take a set of images, place them on the same coordinate frame, and add together. The images must have astrometric information and the quality of the addition will depend on the correction. The result is an exposure time-weighted sum of the inputs. Note that the output pixel is larger than the input pixel, so the counts/pixel will be necessarily larger.\\
\noindent
\textbf{Calling sequence: \textit{jude\_coadd, images\_dir, out\_file, ra\_cent, dec\_cent, fov, pixel\_size}},\\
\noindent 
where {\bit{ra\_cent}} and {\bit{dec\_cent}} are coordinates of the image center, {\bit{fov}} is the image field of view in degrees, and {\bit{pixel\_size}} is in arcsec.

\section{Flux calibration}

The flux calibration of UVIT is described in \cite{Rahna2017}. The flux unit of the final image derived from \jude is counts/second (CPS). CPS can be converted into physical flux in erg s$^{-1}$ cm$^{-2}$ \AA$^{-1}$ by multiplying by the conversion factor $CF$ (erg s$^{-1}$ cm$^{-2}$ \AA$^{-1}$cps$^{-1}$), given in Table~\ref{filterprop} \citep{Rahna2017}:
\begin{equation}
F(\lambda) = CF \times CPS\,.
\label{eq:Fluxconv}
\end{equation}

Errors in CPS can be calculated from the square root of the total number of counts divided by square root of the total exposure time, 
\begin{equation}
CPS_{err}=\frac{\sqrt{CPS}}{\sqrt{exposure~time}}\,.
\label{eq:cpserr}
\end{equation}
To obtain errors in flux $F(\lambda)$, multiply CPS errors by the corresponding conversion factors.

\section{Saturation correction} 

At high count rates, intensified CMOS detectors are subject to non-linearity and saturation. Corrections for saturated bright stars in the UVIT field are discussed in \cite{Rahna2017}, Sec.~3.2.3. We modeled the non-linearity using the formulation of \citet{Kuin2008}. The observed UVIT counts are non-linear after 15 CPS (9.7\% roll-off) and saturated count rates above this value (till 29 CPS) can be corrected using empirical relation,
\begin{equation}
\text{C}_{\rm obs} = 29\times\left(1- \mathrm{e}^{(-\alpha C_{\rm inc}/29)}\right)\,,
\label{eq:non-linearity}
\end{equation}
where $\alpha = 1.24$ (determined empirically), $C_{\rm inc}$ is the number of events incident on the detector, $C_{\rm obs}$ is the number of events detected, and there are 29 frames in a second. After 29 CPS, the counts saturate and the true value cannot be recovered.

\section{Conclusion}

We have described here how to install and operate the independent UVIT data pipeline \jude. The pipeline is free and open-source and consists of modules where users can control and change the parameters. We have successfully used \jude in the last 3 years; first to independently characterize the in-flight performance of the UVIT \citep{Rahna2017} and, secondly, in our scientific programs, of which several are completed \citep{Rahnangc2336, firoza2,firoza1,rubi} and few are still running on \citep[e.g]{rita1, rahnaxuv,yadav}.

\section*{Attribution and Acknowledgements}

If you use this software, please cite \cite{murthy2017jude}. We have used GDL very heavily and here is the attribution for that: \citet{GDL2010}. The IDL Astronomy User's Library reference is \citet{Landsman1995}, the Markwardt IDL Library: \citet{mpfit2009}, and the Astronomy.net reference is  \citet{Astrometry2010}. 

\noindent
The authors gratefully thank all the members of various teams of UVIT project. RPT acknowledge IIA (Indian Institute of Astrophysics, India) and SHAO (Shanghai Astronomical Observatory, China) for providing the computational facilities.

This research has been supported by the Department of Science and Technology under Grant No. EMR/2016/00145 to IIA.

\bibliography{ref}

\end{document}